\pdfoutput=1

\documentclass[9pt]{sig-alternate-05-2015}

\usepackage{amsmath}
\usepackage{graphicx}
\usepackage{booktabs}
\usepackage{amssymb}
\usepackage{latexsym}	
\usepackage{array}		
\usepackage{multirow}
\usepackage{subfigure}
\usepackage{algorithm}
\usepackage{algpseudocode}
\usepackage{threeparttable}
\usepackage{paralist}   
\usepackage{xspace}
\usepackage{color}
\usepackage[table]{xcolor}  
\usepackage{adjustbox}
\usepackage{balance}
\usepackage{wasysym}
\usepackage{rotating}   
\usepackage{listings}
\usepackage{flushend}  
\usepackage{authblk}   
\usepackage{tabu}

\usepackage{url}            
\usepackage[pdfborder={0 0 0}, citecolor=blue, linkcolor=blue, urlcolor=black, colorlinks=true]{hyperref}

\usepackage{xpatch}
\makeatletter
\chardef\TPT@@@asteriskcatcode=\catcode`*
\catcode`*=11
\xpatchcmd{\threeparttable}
  {\TPT@hookin{tabular}}
  {\TPT@hookin{tabular}\TPT@hookin{tabu}}
  {}{}
\catcode`*=\TPT@@@asteriskcatcode
\makeatother

\newcommand{\blue}[1]{\textcolor[rgb]{0.00,0.00,1.00}{#1}}

\definecolor{wheat1}{rgb}{1.000000,0.905882,0.729412}
\definecolor{LightGray}{rgb}{0.827451,0.827451,0.827451}

\newcolumntype{a}{>{\columncolor{wheat1}}l}

\definecolor{mygreen}{rgb}{0,0.6,0}
\definecolor{mygray}{rgb}{0.5,0.5,0.5}
\definecolor{mymauve}{rgb}{0.58,0,0.82}
\definecolor{darkblue}{rgb}{0.0,0.0,0.6}
\definecolor{maroon}{RGB}{102, 0, 0}
\definecolor{Maroon}{cmyk}{0,0.87,0.68,0.32}
\definecolor{darkred}{RGB}{139, 0, 0}
\definecolor{forestgreen}{RGB}{34, 139, 34}

\lstdefinelanguage{XML}
{
  basicstyle=\ttfamily\small,   
  morestring=[b]",
  moredelim=[s][\color{darkblue}]{<}{\ },
  moredelim=[s][\color{darkblue}]{</}{>},
  moredelim=[l][\color{darkblue}]{/>},
  moredelim=[l][\color{darkblue}]{>},
  morecomment=[s]{<?}{?>},
  morecomment=[s]{<!--}{-->},
  stringstyle=\color{darkred},
  identifierstyle=\color{mymauve}
}

\lstdefinestyle{customJava}{
  breaklines=true,
  keepspaces=true,
  frame=single,
  language=Java,
  showstringspaces=false,
  basicstyle=\footnotesize\ttfamily,
  keywordstyle=\color{blue},
  otherkeywords={+, getIntent},
  numbers=left,
  numbersep=5pt,
  numberstyle=\scriptsize\color{black},
  rulecolor=\color{black},
  stepnumber=1,
  tabsize=2,
  commentstyle=\itshape\color{green!40!black},
  stringstyle=\color{orange},
  emph=[1]  
  {
        do,
        try,
        new,
        catch,
        while,
        SecProvider,
        SecReceiver,
        SecService,
        SecActivity,
        SecSink,
  },
  emphstyle=[1]{\color{darkred}},
  emph=[2]  
  {
        @Override,
  },
  emphstyle=[2]{\color{purple!40!black}},
  belowskip=-1em, 
}

\newif\ifANNOYMIZE
\ANNOYMIZEfalse

\newif\ifACM
\ACMtrue  

\ifACM
\newcommand{\myfig}{Figure\xspace}
\else
\newcommand{\myfig}{Fig.\xspace}
\fi

\ifACM
\newcommand{\mysec}{\S}
\else
\newcommand{\mysec}{Section\xspace}
\fi

\newcommand\x{\textcolor[rgb]{1.00,0.00,0.00}{$\times$}\xspace}

\newcommand{\name}{SecComp\xspace}
\newcommand{\entryAC}{\textsc{EP-AC}\xspace}
\newcommand{\sinkAC}{\textsc{SP-AC}\xspace}

\newcommand{\exatk}{EEC\xspace}
\newcommand{\imatk}{IEC\xspace}
\newcommand{\sqlatk}{SQL\xspace}

\makeatletter
\def\@copyrightspace{\relax}
\makeatother

\begin{document}

\title{Where is the Road? Our Vision on Practically Defending Against Component Hijacking in Android Applications}
\title{No More Hijack: Our Vision on Practically Defending Against Component Hijacking in Android Applications}
\title{Our Vision on Practically Defending Against Component Hijacking in Android Applications}
\title{\name: Towards Practically Defending Against Component Hijacking in Android Applications}

\author{
\ifANNOYMIZE
Anonymous Submission
\else
Daoyuan Wu, Debin Gao, Yingjiu Li, and Robert H. Deng\\
\email{\normalsize \{dywu.2015, dbgao, yjli, robertdeng\}@smu.edu.sg}\\
\affaddr{School of Information Systems, Singapore Management University}\\
\affaddr{This is a technical report at \url{http://arxiv.org/abs/1609.03322}, first appeared on 12 September 2016.}
\fi
}

\maketitle

\begin{abstract}

Cross-app collaboration via inter-component communication is a fundamental mechanism on Android.
Although it brings the benefits such as functionality reuse and data sharing, a threat called \textit{component hijacking} is also introduced.
By hijacking a vulnerable component in victim apps, an attack app can escalate its privilege for originally prohibited operations.
Many prior studies have been performed to understand and mitigate this issue, but component hijacking remains a serious open problem in the Android ecosystem due to no effective defense deployed in the wild.


In this paper, we present our vision on practically defending against component hijacking in Android apps.
First, we argue that to fundamentally prevent component hijacking, we need to switch from the previous mindset (i.e., performing system-level control or repackaging vulnerable apps after they are already released) to a more proactive version that aims to help security-inexperienced developers make secure components in the first place.
To this end, we propose to embed into apps a \textit{secure component library} (\name), which performs in-app mandatory access control on behalf of app components.
An important factor for \name to be effective is that we find it is possible to devise a set of practical in-app policies to stop component hijacking.
Furthermore, we allow developers design custom policies, beyond our by-default generic policies, to support more fine-grained access control.
We have overcome challenges to implement a preliminary \name prototype, which stops component hijacking with very low performance overhead.
We hope the future research that fully implements our vision can eventually help real-world apps get rid of component hijacking.

\end{abstract}

\section{Introduction}
\label{sec:intro}

Android has been the dominant player in recent years' smartphone markets.
On Android, different apps are allowed to collaborate with each other via inter-component communication~\cite{UnderstandAndroid09}.
Although such flexible cross-app collaboration brings the benefits such as functionality reuse and data sharing, a threat called \textit{component hijacking}~\cite{CHEX12} is also introduced.
By hijacking a vulnerable component in victim apps, an attack app can bypass Android sandbox and escalate its privilege~\cite{ISC10_Privilege}, causing confused deputy problems~\cite{Confused88} such as permission misuse~\cite{Woodpecker12}, data manipulation~\cite{CHEX12}, and content leaks~\cite{ContentScope13}.

To mitigate component hijacking, many approaches have been proposed.
One major line of the research~\cite{Saint09, IPCInspection11, Quire11, XManDroid11, TrustDroid11, Taming12, IntraComDroid12, SEAndroid13, FlaskDroid13, CICC15} is to modify or extend Android operating system so that inter-component communication could be supervised. 
The other direction~\cite{AppSealer14} is to patch app binaries by performing app repackaging~\cite{Aurasium12, DroidMOSS12}.
Both are useful if they could be deployed in the wild.
But reality is harsh: nearly no proposal\footnote{\small Only SEAndroid~\cite{SEAndroid13} was adopted to replace the Linux UID-based sandbox with the SELinux-confined sandbox~\cite{SELinuxAndroid}.} has been integrated into Android or adopted by Google Play, probably due to the compatibility and performance concerns.
For example, the repackaging approach violates Android's signature-based app verification and thus is not favorable by app markets and developers who own source codes.
Consequently, component hijacking remains a serious open problem in the Android ecosystem.

In this paper, we try to provide a new perspective for the research community to reconsider how to practically defend against component hijacking.
We find that neither earlier proposals consider the problem from developers' perspective;
they are either from the OS or the app market's perspectives.
However, to fundamentally prevent component hijacking, we do need a more proactive mindset that aims to help security-inexperienced developers make secure components in the first place.
To this end, we propose to embed into apps a \textit{secure component library} (\name), which performs in-app mandatory access control on behalf of app components.
Compared to the traditional secure development education, the \name solution is systematic, mandatory, and low-cost.
Moreover, by leveraging automatic code rewriting techniques, we expect that \name could be easily adopted by developers to produce security-enhanced apps which are resistant to component hijacking.

An important enabling factor for \name to work is that we find it is possible to devise a set of practical in-app policies to stop component hijacking.
By analyzing the root causes of different hijacking attacks, we have derived six generic mandatory access control policies for \name to enforce.
In addition, for cases with no fixed patterns, we propose to leverage user-driven access control to handle them.
Furthermore, we allow developers design custom policies, beyond our by-default generic policies, to support more fine-grained access control.

The remainder of this paper is organized as follows.
We first introduce the threat model in \mysec\ref{sec:model}, followed by the design of \name in \mysec\ref{sec:seclib}.
In \mysec\ref{sec:implement}, we present a \name prototype implementation and evaluate it in \mysec\ref{sec:evaluate}.
Related works are outlined in \mysec\ref{sec:related}, and finally we conclude in \mysec\ref{sec:conclude}.

\section{Threat Model}
\label{sec:model}

\begin{figure}[t!]
\vspace{-2ex}
\begin{adjustbox}{center}
\includegraphics[width=0.36\textwidth]{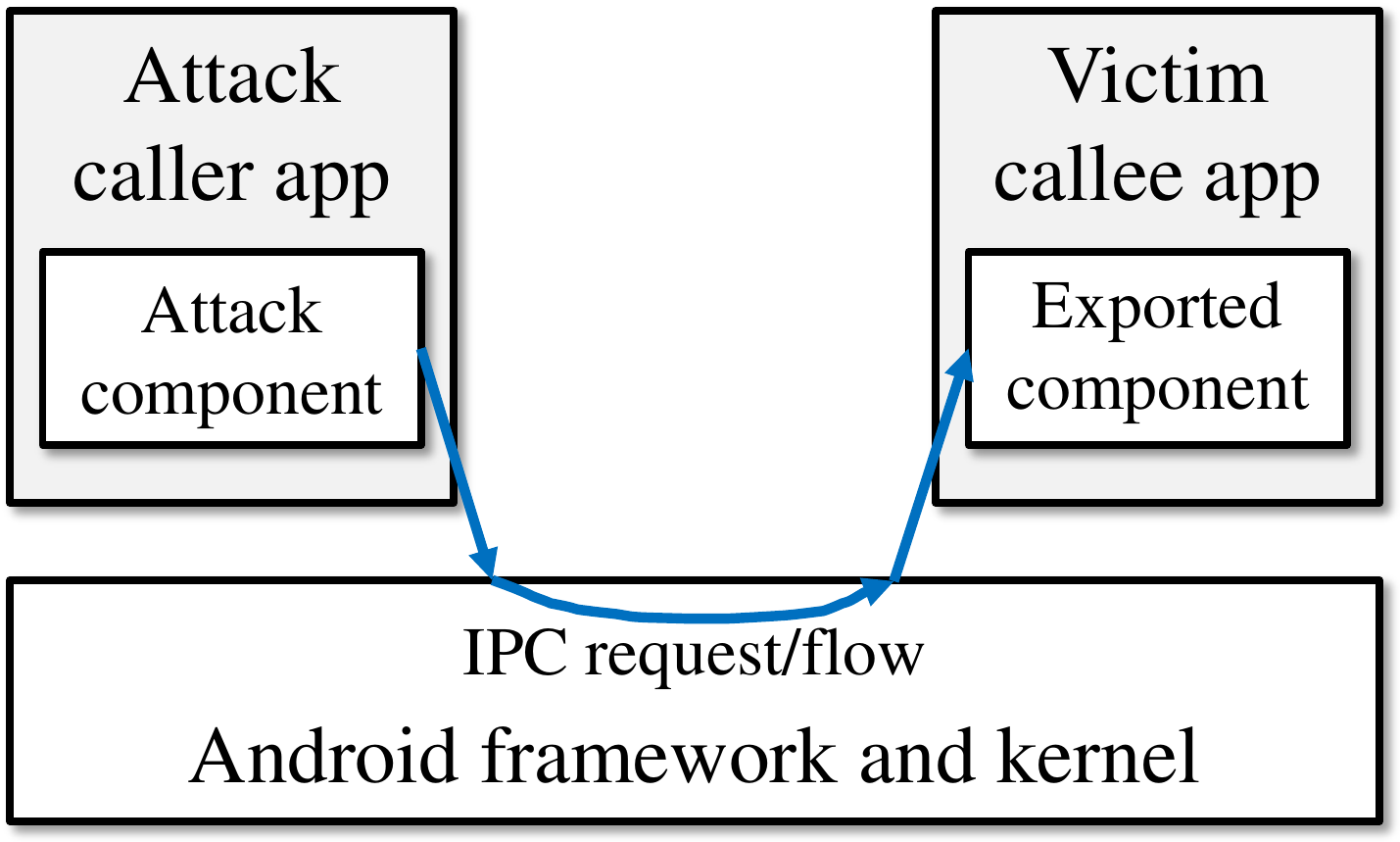}
\end{adjustbox}
\vspace{-5ex}
\caption{\small The threat model of component hijacking.}
\label{fig:threatmodel}
\end{figure}

\myfig~\ref{fig:threatmodel} presents the threat model of component hijacking on Android.
The adversary is a \textit{caller app}, and the victim is a \textit{callee app} that has certain capabilities or owns some sensitive data.
A necessary condition of component hijacking is the target component in the callee app being \textit{exported}, an Android terminology that describes a component is exposed to other apps on the same phone.
The attack component in the caller app then can send a crafted \textit{IPC} (inter-process communication)\footnote{\small In some references, this kind of IPC is also called \textit{ICC} (inter-component communication)~\cite{ICCEpicc13} or inter-app communication. In this paper, we unify these terms using IPC.} request to the exported component, to maliciously trigger its code execution for a privileged operation, e.g., permission misuse~\cite{Woodpecker12} and data manipulation~\cite{CHEX12}.
In this sense, component hijacking belongs to the classic confused deputy problem~\cite{Confused88}.

More specifically, we underline two in-scope threats that are not considered in some related works. 
\begin{compactitem}
\item Unlike~\cite{Woodpecker12, ContentScope13, ECVDetector14}, we do \textit{not} assume that exported components protected with above-\texttt{normal}\footnote{\small There are four levels of Android permissions~\cite{PermissionLevel}: \texttt{normal}, \texttt{dangerous}, \texttt{signature}, and \texttt{signatureOrSystem}.} permissions are always safe.
Because for an exported component protected with an \texttt{dangerous}-level permission, an attack app still can register the corresponding permission for sending IPC requests.
Additionally, a recent report~\cite{CustomPermission} showed that even components with a \texttt{signature}-level permission could be compromised, because the attack app can pre-claim that permission as \texttt{normal} if it is installed earlier than the victim app.

\item Similarly, for the attack app, we do \textit{not} assume it always has zero or few permissions.
Indeed, it can claim the same permission as the misused permission in a victim app.
The benefit for doing so is that it may deceive the IPC call chain-based permission checking~\cite{IPCInspection11, Quire11}.
On the other hand, we \textit{do} assume that the attack app has no root privilege, which is reasonable and a widely adopted practice in Android research.

\end{compactitem}

It is worth noting that a related threat called \textit{unauthorized Intent receipt}~\cite{ComDroid11} is out of the scope of this paper.
This threat is essentially different from component hijacking.
In its model, the attack app is the callee, and vulnerability occurs because the caller victim app mistakenly sends out sensitive information in its IPC messages.
A systematic defense~\cite{Aquifer13} has been proposed to mitigate this issue.

\section{\name: Secure Component Library}
\label{sec:seclib}

\subsection{Overview}
\label{sec:overview}

\begin{figure}[t!]
\vspace{-2ex}
\begin{adjustbox}{center}
\includegraphics[width=0.42\textwidth]{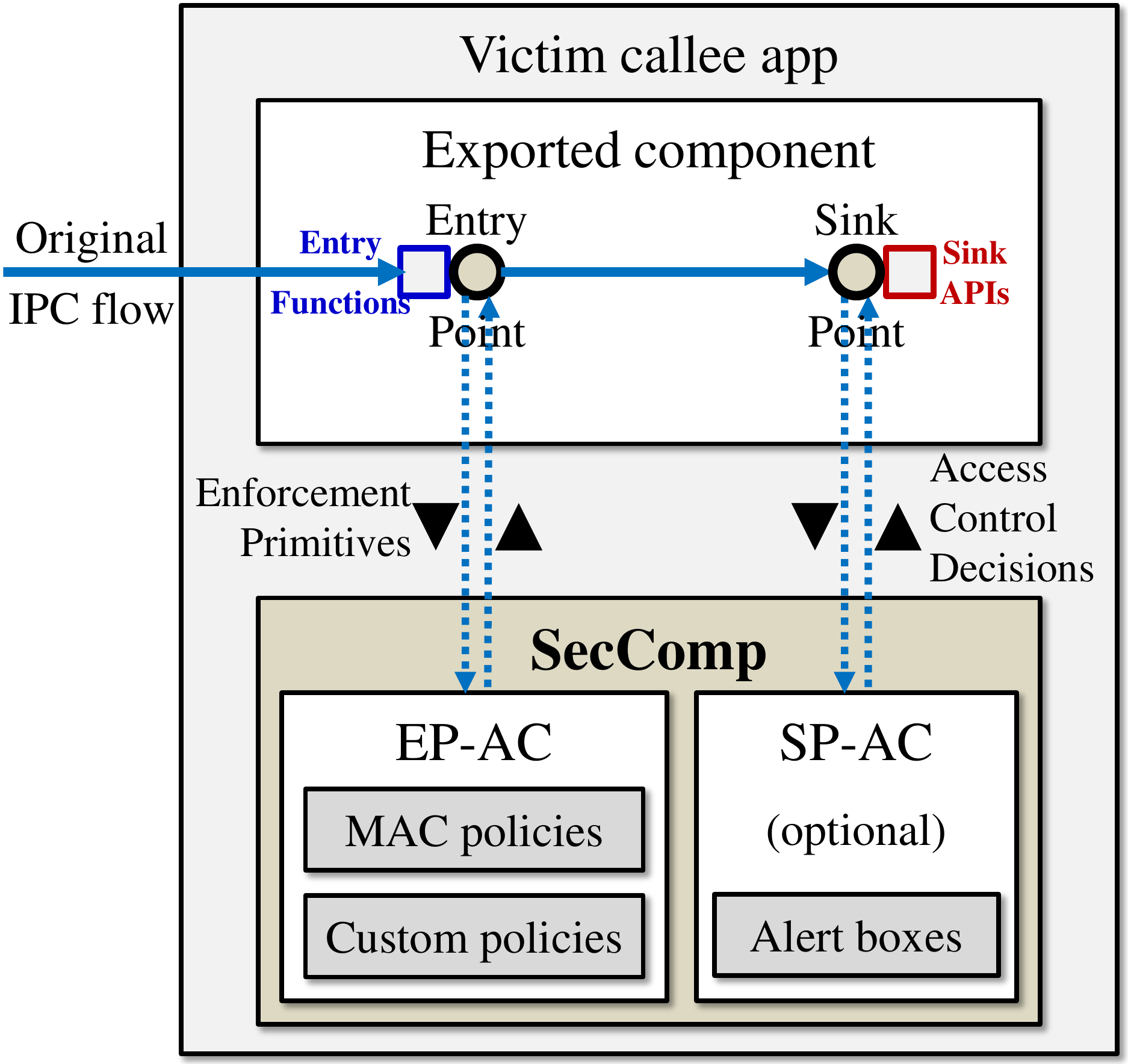}
\end{adjustbox}
\vspace{-5ex}
\caption{\small The overall design of \name. It involves these major designs: (\mysec\ref{sec:entryANDsink}) identifying which entry functions and sink APIs to instrument; (\mysec\ref{sec:instrument}) performing code instrumentation to insert entry and sink points, which route incoming IPC flows to \name for access control; (\mysec\ref{sec:primitive}) extracting enforcement primitives that are used in policy design; and most importantly, (\mysec\ref{sec:policy}) enforcing the mandatory entry-point access control (i.e., \entryAC) and the optional sink-point access control (i.e., \sinkAC). Finally in \mysec\ref{sec:deploy}, we discuss how to deploy \name in the wild.}
\label{fig:seclib}
\vspace{-4ex}
\end{figure}

\myfig~\ref{fig:seclib} presents the overall design of \name, a secure component library for performing in-app mandatory access control on behalf of app components.

\begin{table*}[t!]
\vspace{-2ex}
\centering
\caption{\scriptsize Component entry functions that need to be intercepted, characterized by component types and IPC caller APIs.}
\scalebox{0.9}{
\begin{threeparttable}
\begin{tabu}{ |c | c | l|}

\hline
\rowfont{\bfseries}
\rowcolor{LightGray}
Component types & IPC caller APIs\tnote{\dag} & Interested entry functions\tnote{\dag} of IPC callee components \tabularnewline
\hline
\hline

\multirow{2}{*}{\texttt{Activity}~\cite{Activity}} & \texttt{Context}\cite{Context}\texttt{.startActivity()} & \multirow{2}{*}{\texttt{onCreate(Bundle)}; \texttt{onStart()}; \texttt{onNewIntent(Intent)}}\tabularnewline
& \texttt{Activity.startActivityForResult()} & \tabularnewline
\hline

\multirow{2}{*}{\texttt{Service}~\cite{Service}} & \texttt{Context.startService()} & \texttt{onCreate()}; \texttt{onStartCommand(Intent, int, int)}\tnote{\ddag} \tabularnewline
& \texttt{Context.bindService()} & \texttt{onCreate()}; \texttt{onBind(Intent)}; \texttt{onRebind(Intent)} \tabularnewline
\hline

\multirow{2}{*}{\texttt{BroadcastReceiver}~\cite{BroadcastReceiver}} & \texttt{Context.sendBroadcast()} & \multirow{2}{*}{\texttt{onReceive(Context, Intent)}} \tabularnewline
& \texttt{Context.sendOrderedBroadcast()} & \tabularnewline
\hline

\multirow{6}{*}{\texttt{ContentProvider}~\cite{ContentProvider}} & \texttt{ContentResolver}\cite{ContentResolver}\texttt{.query()}  & \texttt{query(Uri, String[], String, String[], String)} \tabularnewline
& \texttt{ContentResolver.insert()}  & \texttt{insert(Uri, ContentValues)} \tabularnewline
& \texttt{ContentResolver.bulkInsert()}  & \texttt{bulkInsert(Uri, ContentValues[])} \tabularnewline
& \texttt{ContentResolver.update()} & \texttt{update(Uri, ContentValues, String, String[])} \tabularnewline
& \texttt{ContentResolver.delete()}  & \texttt{delete(Uri, String, String[])} \tabularnewline
& \texttt{ContentResolver.openFileDescriptor()}  & \texttt{openFile(Uri, String)} \tabularnewline
\hline

\end{tabu}
\begin{tablenotes}
\item[\dag] For simplicity, we skip the parameters of caller APIs and the class names of entry functions.
\item[\ddag] An old-SDK version of \texttt{onStartCommand(Intent, int, int)} is \texttt{onStart(Intent, int)}, which should be also covered.
\end{tablenotes}
\end{threeparttable}
}
\label{tab:entryfuncs}
\vspace{-2ex}
\end{table*}

\name supports two kinds of enforcements, i.e., entry-point access control (\entryAC) and sink-point access control (\sinkAC).
\entryAC is enforced \textit{just after} the entry of an app component, whereas \sinkAC is performed \textit{just before} each sensitive API call (i.e., the so-called sink, a term commonly used in taint analysis~\cite{FlowDroid14}).
\entryAC serves as the first line of defense, so every incoming IPC request will be checked by this module.
As to be illustrated in \mysec\ref{sec:policy}, we design a set of mandatory access control policies for \entryAC to confine common attacks and mistakes.
Only IPC requests that \entryAC cannot determine legal or not will be further assessed by \sinkAC, at the point a sink API going to be executed.
To guarantee the completeness, we launch user-driven access control in \sinkAC to handle cases with no fixed patterns.

To route incoming IPC flows to \name for access control, we pre-insert entry-point and sink-point instrumentation points into original app component codes.
These instrumentation points are very lightweight---only one or two lines of codes each.
After \name's checking, we either return the execution flow back to the component codes (if the IPC request is legal) or terminate it (if the request is non-legal).
For cases that user decisions are required, we hold on the execution flow and wait for users to make a decision via a pop-up dialog.



\subsection{Identifying Entry Functions and Sink APIs}
\label{sec:entryANDsink}


\noindent\textbf{Entry function identification.}
Table~\ref{tab:entryfuncs} characterizes the component entry functions that need to be intercepted by \name.
We organize them by different component types and IPC caller APIs, and explain them as follows (in the reverse order of component types):

\begin{compactitem}
\item [\texttt{ContentProvider}~\cite{ContentProvider}:]
Identifying entry functions for \texttt{ContentProvider}, the only non-\texttt{Intent} based component among all four types, is straightforward.
Because there is a one-to-one mapping between each caller API and entry function.
For example, if an attack app calls the \texttt{ContentResolver.query} API, the corresponding \texttt{ContentProvider.query} entry function in a victim component will be activated.

\item [\texttt{BroadcastReceiver}~\cite{BroadcastReceiver}:]
There is only one entry for \texttt{BroadcastReceiver}, namely the \texttt{onReceive} callback function.
An adversary can call the \texttt{sendBroadcast} API to trigger \texttt{onReceive}.
Other IPC call APIs that are built upon \texttt{sendBroadcast} could be also used, such as \texttt{sendOrderedBroadcast} and \texttt{sendStickyBroadcast}.
All these calls lead to the execution of \texttt{onReceive} in a callee component.
%

\item [\texttt{Service}~\cite{Service}:]
There are two ways to call a \texttt{Service} component, either by starting it via the \texttt{startService} API, or by binding it via the \texttt{bindService} API.
Both cases will first go through the \texttt{onCreate()} entry function if the service is not created.
The \texttt{onStartCommand} entry (in the first case) operates in a way similar to the \texttt{onReceive} function, whereas the \texttt{onBind} or \texttt{onRebind} entry (in the second case) only returns a \texttt{Binder}~\cite{Binder} object and has no further sequential execution.
Indeed, with the retrieved \texttt{Binder} object, an adversary can invoke any IPC interface functions~\cite{BoundService, AIDL} that are pre-defined by the \texttt{Binder} object.
These custom interface functions thus become additional \texttt{Service} entries, which should be also instrumented.

\item [\texttt{Activity}~\cite{Activity}:]
Similar to \texttt{Service}, the entry functions of \texttt{Activity} also include \texttt{onCreate} and \texttt{onStart}, which could be activated by the \texttt{startActivity()} API and its variants.
Additionally, there is a special entry function in \texttt{Activity}, i.e., \texttt{onNewIntent}.
This entry would be triggered if the caller app sets a special \texttt{Intent} flag called \texttt{FLAG\_ACTIVITY\_SINGLE\_TOP}.
\end{compactitem}
\bigbreak

\noindent\textbf{Sink API selection.}
To pick sink APIs that are relevant to component hijacking, we can leverage some earlier works.
Notably, Stowaway~\cite{Stowaway11} and PScout~\cite{PScout12} released a list of permission-protected APIs, and SuSi~\cite{SuSi14} leveraged supervised learning to output a large set of privacy-related APIs.
These results are very useful, but still need manual efforts to form a hijacking-specific API set.
On the other hand, works for detecting component hijacking issues~\cite{Woodpecker12, CHEX12, ContentScope13} described the sink APIs that are specific to their problems or component types.
Our earlier effort~\cite{ECVDetector14} tried to synthesize these results and has released a preliminary dataset~\cite{VSink}.
In the future, we plan to further collaborate with the open source community to build a comprehensive dataset.


\subsection{Instrumenting Entry and Sink Points}
\label{sec:instrument}

After identifying entry functions and sink APIs, we perform code instrumentation to insert entry and sink points.
To pack the instrumentation codes, \name defines four static Java classes, namely \texttt{SecActivity}, \texttt{SecService}, \texttt{SecReceiver}, and \texttt{SecProvider}.
We find that there are two types of \entryAC instrumentation and one kind of \sinkAC instrumentation:

\begin{lstlisting}[
float=t!,
style=customJava,
basicstyle=\ttfamily\scriptsize,
label={lst:entryreceiver},
caption={\small Instrumenting \texttt{BroadcastReceiver}'s \texttt{onReceive}, an example of the basic \entryAC instrumentation.}
]
public void onReceive(Context cxt, Intent intent) {
+   if (!SecReceiver.receive(cxt, intent))
+       return;
    // Original code
}
\end{lstlisting}

\begin{enumerate}
\item \textbf{Basic \entryAC Instrumentation:}
Most of \entryAC instrumentation just add two lines of codes in the prologue of entry functions.
They intercept original entry parameters and a few context parameters to \name for decision making.
Listing~\ref{lst:entryreceiver} shows the example of instrumenting \texttt{BroadcastReceiver}'s \texttt{onReceive}.
It simply delivers the original \texttt{Context} and \texttt{Intent} parameters to \texttt{SecReceiver}'s \texttt{receive} function for access control (line 2).
If the access is not allowed by policies, the instrumentation code returns without giving the control flow back to original code (line 3).
In most cases, we need to intercept additional parameters such as component \texttt{Context}, from which \name then can extract component identity and attribute for policy enforcement.
To obtain the \texttt{Context}, we deliver a \texttt{this} variable to \entryAC instrumentation.
Moreover, to intercept the incoming \texttt{Intent} for \texttt{Activity}, we invoke the \texttt{getIntent()} API to explicitly obtain its value.

\item \textbf{\entryAC instrumentation for bound services:}
As mentioned earlier, we should treat \texttt{Binder} interface functions as individual entries.
Therefore, for a bound \texttt{Service} such as RemoteService~\cite{AIDL} in Listing~\ref{lst:entryservice}, we instrument not only its \texttt{onBind} as usual, but also the \texttt{getPid} function in a way similar to how we instrument \texttt{ContentProvider}'s entries.

\item \textbf{\sinkAC instrumentation:}
It intercepts both API names and parameters, and delivers them to the user-driven access control module (\mysec\ref{sec:sinkpolicy}) for output.
Listing~\ref{lst:gosms} shows a sample \sinkAC instrumentation for the popular Go SMS Pro app~\cite{GoSmsApp}, which was vulnerable to the \texttt{SEND\_SMS} permission leak~\cite{ComponentHijackingExploit}.
At line 6, \name records the \texttt{sendTextMessage} call and its five parameters, along with the \texttt{this} variable for capturing the component context.
\end{enumerate}

\begin{lstlisting}[
float=t!,
style=customJava,
basicstyle=\ttfamily\scriptsize,
label={lst:entryservice},
caption={\small The special \entryAC instrumentation for a bound \texttt{Service} named RemoteService~\cite{AIDL}. It instruments \texttt{onBind} as usual, and further instruments \texttt{Binder} interface functions (\texttt{getPid} here) as for \texttt{ContentProvider}'s entries.}
]
public class RemoteService extends Service {
    public IBinder onBind(Intent intent) {
+       if (!SecService.bind(this, intent))
+           return null;
        return mBinder;
    }
    private final IRemoteService.Stub mBinder = new IRemoteService.Stub() {
        public int getPid() {
+           if (!SecService.rpc(this,"getPid",null))
+               return null;
            return Process.myPid();
        }
    };
}
\end{lstlisting}

\begin{lstlisting}[
float=t!,
style=customJava,
basicstyle=\ttfamily\scriptsize,
label={lst:gosms},
caption={\small An example of \sinkAC instrumentation for a vulnerable \texttt{Service} component in Go SMS Pro~\cite{GoSmsApp}. The full vulnerablity and exploit codes are available in \cite{ComponentHijackingExploit}.}
]
public void onStart(Intent intent, int startId) {
    String num = intent.getStringExtra("phone");
    String text = getRandomString(8);
    SmsManager SM = SmsManager.getDefault();

+   if (!SecSink.check(this, "sendTextMessage", new Object[]{num, null, text, null, null}))
+       return;
    SM.sendTextMessage(num, null, text, null, null);
}
\end{lstlisting}

\subsection{Extracting Enforcement Primitives}
\label{sec:primitive}

As an in-app defense, \name can collect a set of valuable enforcement primitives that will be used in the policy enforcement:

\begin{compactitem}
\item [\bf{App identity:}]
  We extract the identity information of both caller and callee apps for policy making.
  We use package name (e.g., \texttt{com.whatsapp} for WhatsApp) instead of app UID to represent an app identity, because some apps (e.g., Chrome) may have multiple UIDs.
  We further combine app signatures to determine whether two apps are from the same developer.
  In this paper, we uniformly denote these identity information by $Id_a$ and $Id_v$, which represent the caller and the callee identity, respectively.
  Then, $Id_a \neq Id_v$ represents that an incoming IPC request is from a third-party app.

\item [\bf{Component attribute:}]
  Android apps claim different component attributes in the manifest file, i.e., \texttt{AndroidManifest.xml}.
  We denote these attributes by $XxxAttr$, such as $ExportedAttr$ for the ``exported'' attribute, $PermAttr$ for the permission attribute, and $ActionAttr$ for the \texttt{Intent} actions registered by components.
  Additionally, we denote system-defined permissions and actions by $SysPerms$ and $SysActions$, respectively.
  A list of permissions and actions that are defined by Android is available in~\cite{SysPerms, SysActions}.
  If a permission can be claimed only by system or an action can be sent only by system, we denote them by $SysOnlyPerm$ and $SysOnlyAction$, respectively.

\item [\bf{Entry-point input data:}]
  Likewise, we denote different input values by $InputXxx$.
  For example, the incoming \texttt{Intent} action is represented as $InputAction$, and the incoming data \texttt{Uri} is denoted by $InputUri$.

\item [\bf{Sink-point parameters:}]
  We denote a sink API parameter by $SinkPara$.
  This primitive is used only by \sinkAC.
\end{compactitem}

In particular, we will detail how to collect app identity and component attributes in \mysec\ref{sec:implement2}. 
The input data and sink parameters can be obtained through the entry-point and sink-point instrumentation codes (see Listing~\ref{lst:entryreceiver} and~\ref{lst:gosms}).

\subsection{Enforcing In-App Access Control}
\label{sec:policy}

\begin{table*}[ht!]
\vspace{-2ex}
\centering
\caption{\small MAC policies for \entryAC. Here we list six representative policies (P1 to P6) we have designed.}
\scalebox{0.9}{
\begin{threeparttable}
\begin{tabu}{ |c | l | c | l|}

\hline
\rowfont{\bfseries}
\rowcolor{LightGray}
ID & Policy Name & \dag & Policy Representation \tabularnewline
\hline
\hline

P1 & No Pre-claimed Custom Permission
& All
& \textbf{if} \blue{$Id_{a} \neq Id_{v}$} $\land$ \blue{$\exists(PermAttr_{v} \notin SysPerms)$} $\land$ \blue{$PermAttr_{v} = PermAttr_{a}$}: \textbf{deny}
\tabularnewline \hline

P2 & No By-default Exported Provider
& P
& \textbf{if} \blue{$Id_{a} \neq Id_{v}$} $\land$ \blue{$\neg ExportedAttr$}: \textbf{deny}
\tabularnewline \hline

\multirow{2}{*}{P3} & No By-default Exported Component
& \multirow{2}{*}{A,S,B}
& \multirow{2}{*}{\textbf{if} \blue{
  $Id_{a} \neq Id_{v}$} $\land$ \blue{$\neg ExportedAttr$} $\land$ \blue{$ActionAttr \notin SysActions$
  }: \textbf{deny}}
\tabularnewline
& for Custom Intent Action & & \tabularnewline
\hline

P4 & Checking System-only Broadcasts
& B
& \textbf{if} \blue{$Id_{a} \neq Id_{v}$} $\land$ \blue{$\exists (ActionAttr \in SysActions)$} $\land$ \blue{$InputAction \neq ActionAttr$}: \textbf{deny}
\tabularnewline \hline

P5 & Stopping DoS (Denial-of-Service)
& All
& \textbf{if} \blue{$Id_{a} \neq Id_{v}$} $\land$ \blue{$CrashNum \geq Threshold$}: \textbf{deny}
\tabularnewline \hline

P6 & Filtering Sql Injection for Provider
& P
& \textbf{if} \blue{$Id_{a} \neq Id_{v}$} $\land$ \blue{$\exists (AttackStr \in InputPara)$}: \textbf{deny}
\tabularnewline \hline

\end{tabu}
\begin{tablenotes}
\item[\dag] This column lists which components this policy is applicable for. The five symbols are explained as follows.\\
  All: all four components; A: \texttt{Activity}; S: \texttt{Service}; B: \texttt{BroadcastReceiver}; P: \texttt{ContentProvider}.
\end{tablenotes}
\end{threeparttable}
}
\label{tab:macpolicy}
\vspace{-2ex}
\end{table*}

\subsubsection{MAC Policies for \entryAC}
\label{sec:macpolicy}

We find that it is possible to devise mandatory access control (MAC) policies to stop common component hijacking issues.
These issues arise because of system flaws or developer mistakes.
We study their root causes and design the corresponding MAC policies for \entryAC to enforce.
Table~\ref{tab:macpolicy} lists six representative MAC policies (P1 to P6) we have designed.
From a high-level view, policies P1 to P3 patch the system weaknesses, P4 and P5 mitigate common developer mistakes, and P6 filters a common attack.

A common point among all six policies is that we treat the IPC requests initiated from the same app or developer trusted.
That is, only an external IPC request from a third-party app will be checked.
This is denoted by $Id_{a} \neq Id_{v}$, as mentioned in \mysec\ref{sec:primitive}.
However, an experienced adversary may bypass this policy by detouring IPC requests first to another component in the victim app.
We propose to mitigate this problem by adding a flag during each IPC relay so that \name can infer the origin of an IPC call chain.
For example, we can add one line of code, \texttt{intent.putExtra(`seclibflag', `outside')}, into each \sinkAC instrumentation that contains IPC call functions.

The policy \textbf{P1} is derived from a system flaw~\cite{CustomPermission} that an attack app who installed earlier can pre-claim a custom permission in the victim app, such as setting its permission level from \texttt{signature} to \texttt{normal}.
Consequently, the attack app can hijack any component that was originally protected with \texttt{signature}-level permissions.
Based on this root cause, our policy first determines whether there is a custom permission defined in the callee component, i.e., $\exists(PermAttr_{v} \notin SysPerms)$.
If there is a such custom permission, we further check whether or not it has been pre-claimed by the caller app, i.e., $PermAttr_{v} = PermAttr_{a}$.
If yes, we deny the request to prevent a component hijacking attack.

The policy \textbf{P2} and \textbf{P3} aim to mitigate the gap that components could be by default exported by system whereas developers may not be aware of that.
Specifically, components that define Intent Filters are implicitly exported~\cite{Intent} even if they do not claim the ``exported'' attribute.
Furthermore, \texttt{ContentProvider} components are automatically exported unless developers explicitly assign ``false'' to the ``exported'' attribute.
This by-default rule led to thousands of vulnerable \texttt{ContentProvider} components~\cite{ContentScope13}.
Although Google later revised this rule since Android 4.2, it still requires developers to manually update apps' SDK attributes to a safe version.
According to a very recent study~\cite{TargetFragmentation16}, there are still many by-exported \texttt{ContentProvider} components.

To protect by-default exported component from hijacking, policy \textbf{P2} and \textbf{P3} first determine whether there exists the ``exported'' attribute.
For \texttt{ContentProvider}, we can deny the IPC request if there is no ``exported'' attribute, in order to mimic the current system rule.
For other components, we further check what kinds of \texttt{Intent} actions are registered.
Only for those custom actions ($ActionAttr \notin SysActions$), the corresponding IPC requests will be denied.
Additionally, we will provide a debug mode with user interfaces, to help developers resolve the potential (but less likely) conflict between their original intentions and our policies.

Following P3, we further propose the policy \textbf{P4} to handle the case of system actions.
In particular, a component listening to system-only broadcasts is hijack-able if it does not check the incoming \texttt{Intent} action explicitly in the code.
A prior work \cite{ECVDetector14} showed that there are around 150 system-only broadcasts on Android.
To prevent component hijacking due to missed action checks, P4 automatically checks the input action against the system-only action claimed in manifest, i.e., $\exists (ActionAttr \in SysAction)$ $\land$ $InputAction \neq ActionAttr$.

The policy \textbf{P5} aims to mitigate the denial-of-service hijacking due to missed null checks on IPC input.
This is a common mistake that affected a large portion of Android apps, according to a previous empirical study~\cite{Study11}.
To stop such hijacking, we record the app crash times corresponding to each caller app.
And if it has exceeded the threshold value (e.g., 3 times), \name then denies the request. 
Moreover, we can consider only the recent crashes to avoid over-checking.

Finally, we propose the policy \textbf{P6} to filter SQL injection for \texttt{ContentProvider}.
As demonstrated in~\cite{ContentScope13}, an attack app can hijack a provider component to inject malicious SQL statements.
For example, the adversary sets the \texttt{projection} parameter of the \texttt{query} function as a special phase ``\texttt{* from private\_table;}''.
Since these special inputs are usually different from normal queries, we thus can use keyword-based filtering (such as the keyword ``\texttt{;}'') to stop them.

\subsubsection{User-driven Control for \sinkAC}
\label{sec:sinkpolicy}

\begin{figure}[t!]
\begin{adjustbox}{center}
\includegraphics[width=0.18\textwidth]{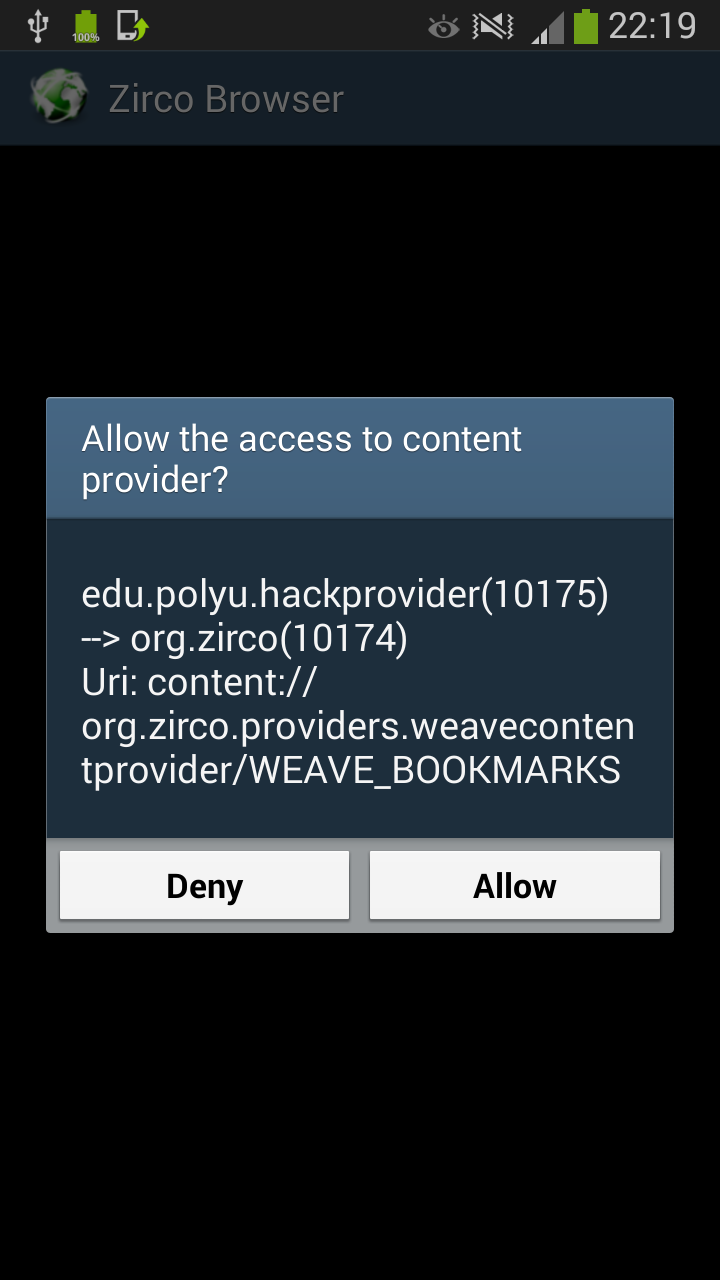}
\end{adjustbox}
\caption{\small An UI example of user-driven control.}
\label{fig:alertbox}
\end{figure}

For IPC requests that cannot be determined by \entryAC, we further launch the user-driven control when they are going to arrive at a sink point.
Basically, \name pops out an alert dialog interface and asks users to make a decision.
\myfig~\ref{fig:alertbox} shows a sample alert box that we have implemented.
By collecting all the enforcement primitives in \mysec\ref{sec:primitive}, we can provide sufficient context information to help users choose ``Deny'' or ``Allow''.
We plan to also leverage recent advances on usable security~\cite{Remystify15, BrowWarn13, UDAC12} to make \name's user-driven control more accessible to end users.
Furthermore, we allow developers to revise each individual pop-up dialog through \name's debugging mode. 
They can even draft custom policies to black- or white-list certain sink points.

It is also possible to derive MAC policies for \sinkAC, although we currently use the user-driven control scheme.
One potential MAC policy is to deny all background sink access (initiated from a third-party app) when users are not using their phones.
The policy P6 in Table~\ref{tab:macpolicy} could be also evolved to a sink-point version, i.e., preventing SQL injection for database sink APIs in all components.
Moreover, we may leverage the IPC call chain-based permission checking~\cite{IPCInspection11, Quire11} to perform mandatory access control.
For instance, if the caller app has no \texttt{SEND\_SMS} permission, it should be also denied to access a component that has such permission.
In practice, this kind of MAC needs more careful design to minimize the negative impact to normal functionalities.
For example, it is legal for a file management app (with zero permission) to instruct the Dropbox app to upload a file.

\subsubsection{Towards Fine-grained Custom Policies}
\label{sec:custompolicy}

As both \entryAC and \sinkAC policies are general-purpose and made by us, there is a need for individual developer to craft app-specific policies for more fine-grained access control. 
We plan to support such custom policies in \name, which enables new use cases.
We introduce two cases below: 
\begin{compactitem}
\item [\textit{Whitelist and blacklist:}]
  Developers can write custom policies to support whitelist and blacklist based access control. 
  For example, they can specify \textit{App A is allowed to access my Component C}.
  This is in particular useful to big vendors that own several development teams with each using different app signature.
  By using whitelist, apps made by different teams can still easily collaborate with one another.

\item [\textit{Policy update and hotfix:}] 
  Using custom policies allows developers to update their policies on-the-fly, such as pushing hotfixes to protect their apps from being exploited by a newly identified system flaw.
\end{compactitem}

\subsection{Deploying \name in the Wild}
\label{sec:deploy}

The lightweight nature and requiring no OS change make \name have a great potential to be deployed in the wild.
The primary \name users are app developers.
They are free to choose any of the following two deployment models:

\begin{compactitem}
\item [\textit{Selective \name planting:}]
Developers can selectively plant \name to protect particular components that are more likely to be attacked.
Specifically, developers import the \name jar file as like using other third-party libraries.
Once imported, they can selectively add the instrumentation codes as in \mysec\ref{sec:instrument}.

\item [\textit{Automatic \name planting:}] 
To further minimize developers' efforts, we are on the way of implementing a code rewriting technique to automatically plant \name.
It can be a standalone script tool or a plugin integrated in Eclipse and Android Studio.
The basic workflow is like this.
By default, it locates the entry functions of each exported component and inserts the \entryAC codes.
Optionally, developers could further instruct this tool to perform control flow analysis to identify reachable sink APIs and adds \sinkAC.
\end{compactitem}

Besides app developers, end users can directly use \name by incorporating the recent advance of app sandboxing called Boxify~\cite{Boxify15}.

%
%

\section{Implementing A \name Prototype}
\label{sec:implement}

We have implemented a \name prototype for \texttt{ContentProvider} components, i.e., \texttt{SecProvider}.
We choose \texttt{ContentProvider} as the first implementation choice because only this type of components has no known defense proposal.
Implementing \name for other three components is generally similar and also underway.
In the course of implementing \texttt{SecProvider}, we have identified and overcame two main challenges:
(\mysec\ref{sec:implement1}) how to achieve user-driven control for background components such as \texttt{ContentProvider};
(\mysec\ref{sec:implement2}) how to extract app identity and component attributes within the context of different Android components.
In particular, we have developed a novel UI (user interface) transition technique to overcome the first challenge.

\subsection{Achieving User-driven Access Control}
\label{sec:implement1}

To achieve user-driven access control, we need to pop out an alert box as in \myfig \ref{fig:alertbox}.
However, we find that this is a non-trivial task for \texttt{ContentProvider} components.
Because \texttt{ContentProvider} is the only type of components that retrieves a caller app's \texttt{Context} object instead of a callee app's~\cite{Context}.
As it is forbidden to display one app's UI elements using another app's \texttt{Context}, \name cannot properly pop out an alert box for \texttt{ContentProvider}.

We take a different strategy to address this problem.
Instead of directly displaying an alert box on top of a caller app, we initialize a dialog-like \texttt{Activity} and use the caller app's \texttt{Context} object to launch this alert \texttt{Activity} via the \texttt{startActivity()} API.
It is worth noting that launching \texttt{Activity} components from \texttt{ContentProvider} requires to set a special \texttt{Intent} flag called \texttt{FLAG\_ACTIVITY\_NEW\_TASK}.
Besides popping out a dialog-like \texttt{Activity}, we also need a method to automatically return to the original caller UI after users click alert box buttons.
This is quite important for maintaining user experience.
A trivial solution is to invoke the \texttt{startActivity} again with the caller component as the target.
We find a more efficient way.
By setting one more special \texttt{Intent} flag called \texttt{FLAG\_ACTIVITY\_MULTIPLE\_TASK}, our alert \texttt{Activity} can naturally go back to the caller UI after invoking the \texttt{Activity.finish()} API.
Moreover, we set the third \texttt{Intent} flag called \texttt{FLAG\_ACTIVITY\_EXCLUDE\_FROM\_RECENTS} to avoid the alert box appear in the history of recent \texttt{Activity} components.

After successfully launching the alert dialog in \texttt{ContentProvider}, we need a mechanism to pause the current component execution and wait for users' decisions.
To do so, \name initializes a lock object and sets this object into the ``wait'' status after invoking the \texttt{startActivity}.
Once users click the buttons, the alert \texttt{Activity} will notify the ``waited'' lock object to be released.
Consequently, the lock object will no longer wait and the paused \texttt{ContentProvider} component resumes its execution.
To avoid the unnecessary long-time waiting, we further set a timeout value (e.g., 30 seconds in our current prototype).
Even if users do not make decisions after the timeout value, \name will by default deny the access.

\subsection{Extracting App Identity and Component Attributes}
\label{sec:implement2}

We now present the implementation of the enforcement primitive extraction as previously shown in \mysec\ref{sec:primitive}.
Among the four types of primitives, obtaining entry-point input data and sink-point parameters are straightforward via our instrumentation code in Listing~\ref{lst:entryreceiver} and \ref{lst:gosms}.
Therefore, we focus on how to extract app identity and component attributes.

\textbf{App identity extraction.}
To extract the package name of a caller app, we can use the \texttt{getCallingPackage()} API.
However, this API is only available since Android 4.4 for \texttt{ContentProvider}.
To maintain the compatibility of \name, we opt to a generic way.
We first extract the caller UID via \texttt{Binder.getCallingUid()} and then use the UID to retrieve the caller package.
Meantime, the callee app package can be directly retrieved from \texttt{ApplicationInfo}, which is available through \texttt{Context.getApplicationInfo()}.

\textbf{Component attribute extraction.}
To illustrate component attribute extraction, we use the $ExportedAttr$ as a representative example.
It aims to determine the export status of a callee component, i.e., explicitly or implicitly exported.
Unfortunately, this is non-trivial.
Initially, we try to obtain the status via the \texttt{ProviderInfo.exported} flag.
But we find that Android has already set this flag to true even for those implicitly exported components.
Consequently, it is impossible to determine the real export status according to the \texttt{exported} flag.

Without the Android API support, we have to seek another way to obtain the component export status.
Our solution is to retrieve the whole \texttt{AndroidManifest.xml} file of the callee app and parse the export status from that.
More specifically, if our parser does not encounter the \texttt{exported} attribute, we determine the corresponding component is implicitly exported.

\section{Evaluation}
\label{sec:evaluate}

\subsection{Security Evaluation}
We now evaluate the effectiveness of \name against different hijacking attacks in \texttt{ContentProvider}.
We use the open source Zirco Browser~\cite{zirco} as our evaluation subject.
It has the following two \texttt{ContentProvider} components, the explicitly exported \texttt{WeaveContentProvider} and the implicitly exported \texttt{ZircoBookmarksContentProvider}.

\begin{lstlisting}[
language=Xml,
basicstyle=\ttfamily\small,
label={lst:manifest},
caption={\small The two exported \texttt{ContentProvider} components in Zirco Browser.}
]
<provider android:name=".providers.WeaveContentProvider" android:authorities="org.zirco.providers.weavecontentprovider" android:exported="true"/>
<provider android:name=".providers.ZircoBookmarksContentProvider" android:authorities="org.zirco.providers.zircobookmarkscontentprovider"/>
\end{lstlisting}

We have implemented \name to defend against three \texttt{ContentProvider} hijacking attacks:
(i) \textbf{the \exatk attack} that exploits explicitly exported components,
(ii) \textbf{the \imatk attack} that exploits implicitly exported components,
and (iii) \textbf{the \sqlatk attack} that performs SQL injections at \texttt{ContentProvider} components.
More specifically, we leverage the policy P2 and P6 (in Table~\ref{tab:macpolicy}) to stop the last two attacks, and further extend the policy P2 as the following custom policy to defend against the \exatk attack.
\begin{equation*}
\textbf{if} \blue{ Id_{a} \neq Id_{v}} \land \blue{ExportedAttr = true}: \textbf{alert}
\label{equ2}
\end{equation*}

\begin{figure}[t!]
\begin{adjustbox}{center}
  \subfigure[No \name.] {
    \label{fig:defendEEC1}
    \includegraphics[width=0.235\textwidth]{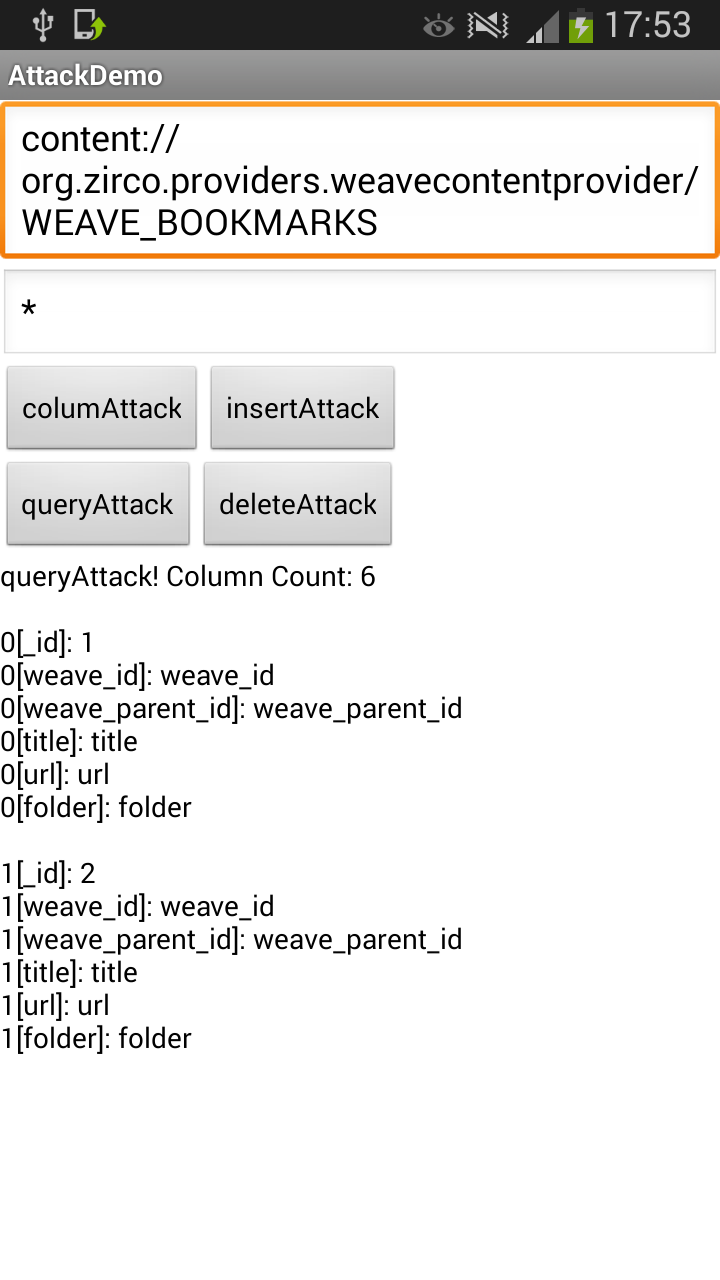}
  }
  \subfigure[With \name.] {
    \label{fig:defendEEC2}
    \includegraphics[width=0.235\textwidth]{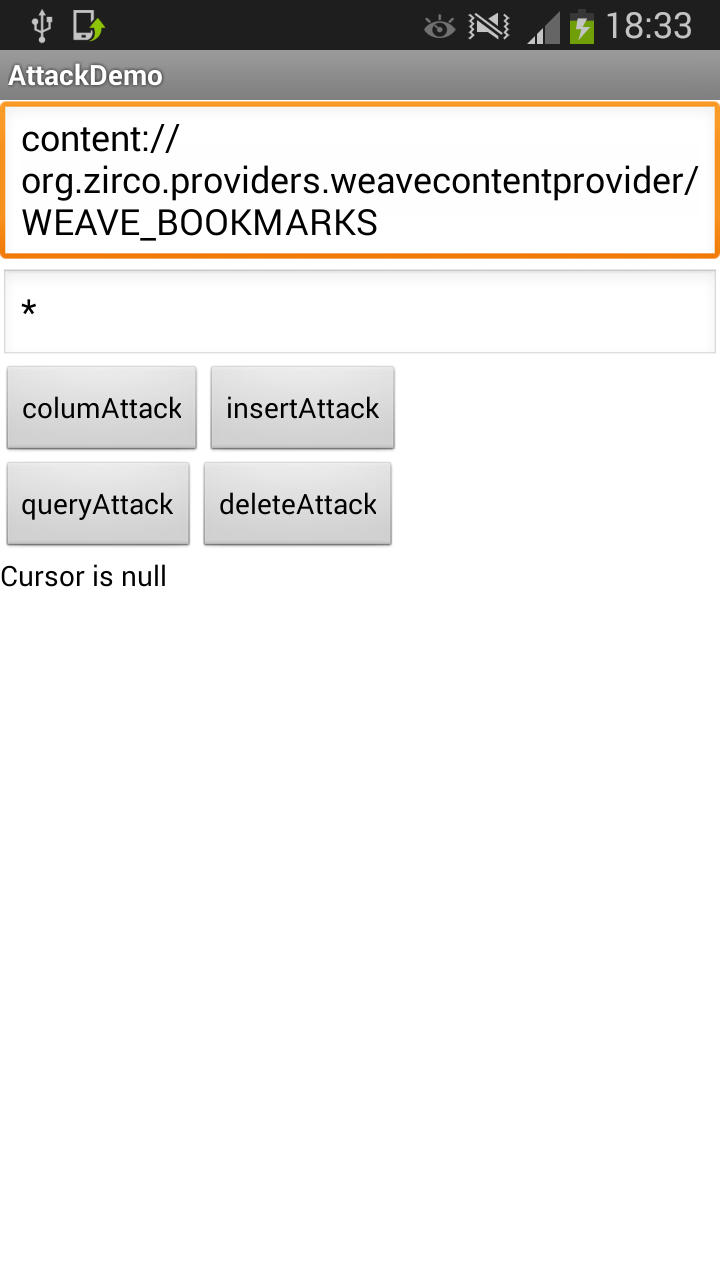}
  }
\end{adjustbox}
\caption{\small Defending the \exatk attack.}
\label{fig:defendEEC}
\end{figure}

\textbf{Defending the \exatk attack.}
We use \texttt{WeaveContentProvider} to evaluate \name against the \exatk attack.
As shown in \myfig~\ref{fig:defendEEC1}, without \name, an adversary can access all information from the \texttt{weave.db} database when our library is not included into the Zirco Browser.
After \name is included in the Zirco Browser, it will pop out an alert dialog for users to do control (see \myfig~\ref{fig:alertbox}). 
When users select the ``Deny'' button, no information could be accessed from the \texttt{WeaveContentProvider}, as shown in \myfig~\ref{fig:defendEEC2}.

\begin{figure}[t!]
\begin{adjustbox}{center}
  \subfigure[No \name.] {
    \label{fig:defendIEC1}
    \includegraphics[width=0.235\textwidth]{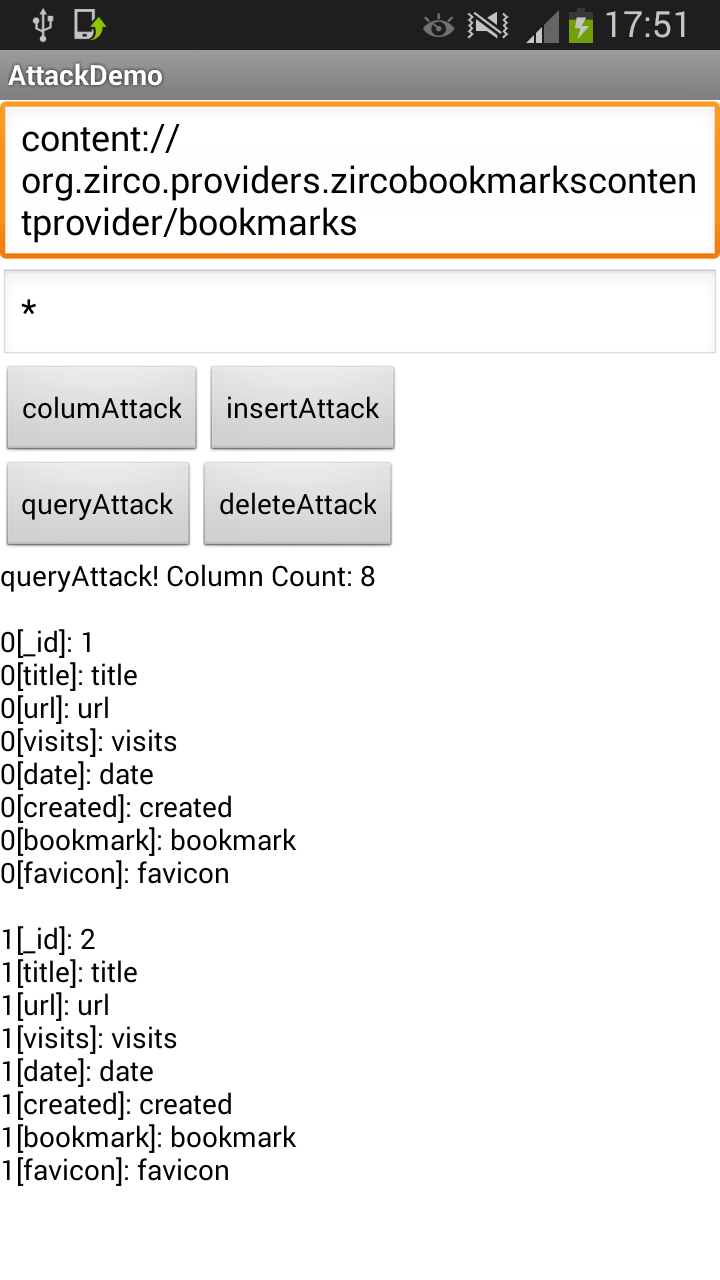}
  }
  \subfigure[With \name.] {
    \label{fig:defendIEC2}
    \includegraphics[width=0.235\textwidth]{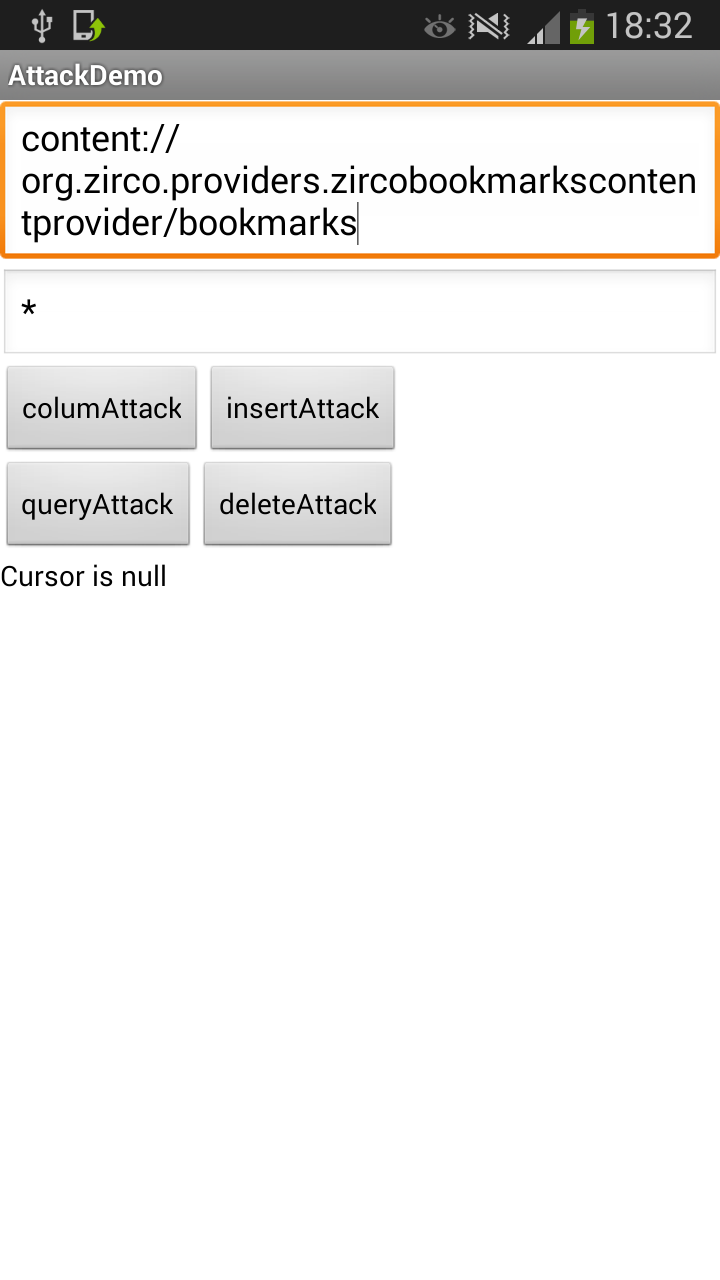}
  }
\end{adjustbox}
\caption{\small Defending the \imatk attack.}
\label{fig:defendIEC}
\end{figure}

\textbf{Defending the \imatk attack.}
We use \texttt{ZircoBookmarksContentProvider} to evaluate \name against the \imatk attack.
As shown in \myfig \ref{fig:defendIEC1}, an adversary can access all information from the \texttt{bookmarks.db} database when \name is not included into the Zirco Browser.
After including \name, as shown in \myfig \ref{fig:defendIEC2}, it will directly deny IPC requests from other apps. 
As a result, no information of \texttt{ZircoBookmarksContentProvider} could be leaked any more.

\begin{figure}[t!]
\begin{adjustbox}{center}
\includegraphics[width=0.235\textwidth]{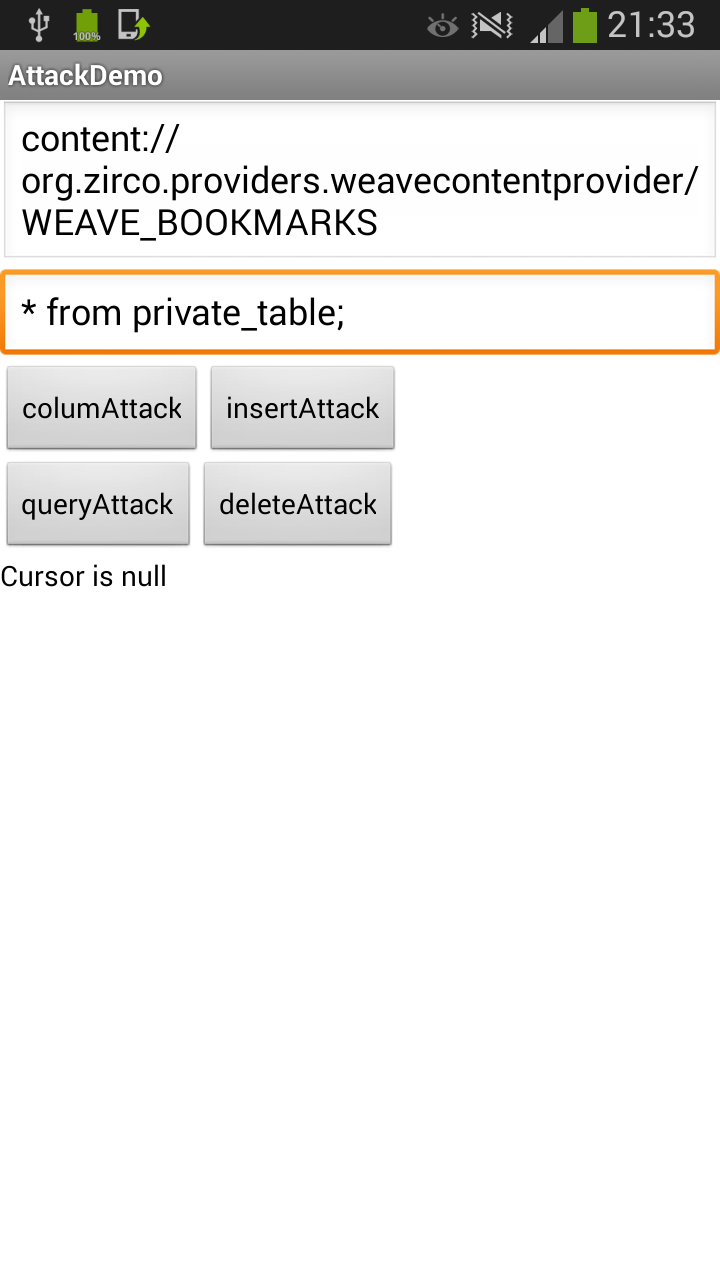}
\end{adjustbox}
\caption{\small Defending the \sqlatk attack.}
\label{fig:defendSQL}
\end{figure}

\textbf{Defending the \sqlatk attack.}
We still use \texttt{WeaveContentProvider} to evaluate \name against the \sqlatk attack.
This time the adversary sets the \texttt{projection} parameter as a special phase ``\texttt{* from private\_table}'', which can launch the SQL injection.
But \name can directly defend against such malicious behaviors, as shown in \myfig \ref{fig:defendSQL}.

\textbf{Self protection.}
The adversary may exploit the fact that \name for stopping the \exatk attack pops out an alert dialog to launch the GUI-based denial-of-service attack.
To prevent such attack, we can provide users a choice button called ``Remember my decision'' so that next time no alert dialog will be popped out.
We also can set a threshold value that defines how many alert dialogs could be popped out within a time slot.

\subsection{Performance Evaluation}

We further evaluate the overhead introduced by \name.
To this end, we record the execution time of \name under different operations.
It is worth noting that for measuring the overhead of defending the \exatk attack, we temporarily remove the lock wait code.
Because user decision time should not be included as our overhead.

Table \ref{tab:overhead} shows the overhead results.
We can see that the overhead of defending the \sqlatk attack is the smallest, only requiring 0.008--0.010ms with the 95\% confidence interval. 
This is because \name only needs to check whether input data contains malicious strings for defending the \sqlatk attack.
The overhead of normal component operation is the second smallest, requiring only 0.235--0.265ms.
This is because we only add one more checking at this step, i.e., to check the caller app identity.
We further add the component semantic checking to defend against the \imatk attack, the overhead of which is 0.243--0.369ms.
As the overhead of all these three operations is under 0.5ms, we conclude that their overhead is very low.
The overhead of defending \exatk attack is the largest, requiring around 200ms, because launching an alert \texttt{Activity} is a bit expensive.
But since it is finished within 0.5s, we believe that this is still a low overhead.

\begin{table}[t!]
\caption{\scriptsize The overhead introduced by \name.}
\begin{adjustbox}{center}
\scalebox{0.9}{
\begin{tabular}{ |c | c|}

\hline
 & Overhead \tabularnewline
 & (mean, 95\% confidence interval) \tabularnewline
\hline
\hline

Defending the \sqlatk attack & $0.009 \pm 0.001$ ms  \tabularnewline
\hline
Normal component operation & $0.250 \pm 0.015$ ms \tabularnewline
\hline
Defending the \imatk attack & $0.306 \pm 0.063$ ms \tabularnewline
\hline
Defending the \exatk attack & $229.417 \pm 45.336$ ms \tabularnewline
\hline

\end{tabular}
}
\end{adjustbox}
\label{tab:overhead}
\end{table}

\section{Related Work}
\label{sec:related}

Many works have been conducted to mitigate component hijacking on Android or secure cross-app communication in general.
Most of them~\cite{Saint09, IPCInspection11, Quire11, XManDroid11, TrustDroid11, Taming12, IntraComDroid12, SEAndroid13, FlaskDroid13, ASF14, CICC15, DIFC16} aim to evolve Android's security architecture by introducing fine-grained, mandatory, and extensible access control at different system levels.

A notable example is SEAndroid~\cite{SEAndroid13}, which reconstructs Android from the previous Linux UID-based discretionary access control to the recent SELinux-enabled mandatory access control.
It can restrict certain app flaws such as direct file leak~\cite{MoST15, FileCross14} but not component hijacking, because it is quite challenging to efficiently audit every IPC at the system level without affecting any normal app functionality. 
Probably due to the same reason, Android IntentFirewall is still an experimental feature~\cite{IntentFirewallHomepage, IntentFirewallDoc} despite its code release over three years ago~\cite{IntentFirewallCode}.
Moreover, IntentFirewall is a \textit{sender}-oriented policy framework~\cite{IntentScope16} and thus requires the malicious Intent patterns.  
In contrast, \name has no such requirement and can be applied to both Intent and non-Intent based components.
Additionally, Kantola et. al.~\cite{IntraComDroid12} made OS change to restrict component exposure.
Although this is effective to reduce attack surfaces, it is not flexible and cannot handle some component hijacking attacks such as those defended by our policy P1, P4, P5, and P6.

The other line of defense solutions is to perform app repackaging~\cite{Aurasium12, DroidMOSS12, PushPopRandom15}.
Notably, AppSealer~\cite{AppSealer14} repackages apps to generate component hijacking patches which performs the sink-point enforcement. 
Aurasium~\cite{Aurasium12} offers a similar capability although it is not specialized for component hijacking.
Our \sinkAC is inspired by these works but not in a form of repackaging.

As orthogonal to the defense research, many prior studies try to understand and detect component hijacking issues in real-world apps.
They have leveraged program analysis techniques to propose various detection systems, including ComDroid~\cite{ComDroid11}, Woodpecker~\cite{Woodpecker12}, CHEX~\cite{CHEX12}, ContentScope~\cite{ContentScope13}, and Epicc~\cite{ICCEpicc13}.
Android app analysis frameworks, such as FlowDroid~\cite{FlowDroid14} and Amandroid~\cite{Amandroid14}, could be also extended to analyze component hijacking.
Recently, more solid and scalable inter-component analysis methods~\cite{IC315, MarketICC16} have also been proposed.

\section{Conclusion and Future Works}
\label{sec:conclude}

In this paper, we presented our vision on practically defending against component hijacking in Android apps.
We proposed \name, a secure component library for performing in-app mandatory access control on behalf of app components.
We further designed a set of generic and practical policies for \name to enforce.
We have implemented a preliminary \name prototype and demonstrated its efficacy and efficiency.
We are on the way of implementing a complete \name system, and will open source it soon so that more hands could join to help real-world apps get rid of component hijacking.


{\small \bibliographystyle{acm}
\bibliography{main}}

\end{document}